\documentclass[10pt,msam,msbm]{iopart}
\usepackage{iopams}
\usepackage{graphicx}
\usepackage{dcolumn}
\usepackage{bm}
\usepackage{color}

\begin{document}

\title[]{Probing superconductivity in MgB$_{2}$ confined to magnetic field tuned cylinders by means of critical fluctuations}

\author{S~Weyeneth$^1$, T~Schneider$^1$, N~D~Zhigadlo$^2$, J~Karpinski$^2$ and H~Keller$^1$}

\address{$^1$~Physik-Institut der Universit\"{a}t Z\"{u}rich, Winterthurerstrasse 190, CH-8057 Z\"urich, Switzerland}
\address{$^2$~Laboratory for Solid State Physics, ETH Z\"urich, CH-8093 Z\"{u}rich, Switzerland}

\ead{wstephen@physik.uzh.ch}
\begin{abstract}
We report and analyze reversible magnetization measurements on a
high quality MgB$_{2}$ single crystal in the vicinity of the zero
field transition temperature, $T_{c}\simeq 38.83$ K, at several
magnetic fields up to $300$ Oe, applied along the $c$-axis. Though
MgB$_{2}$ is a two gap superconductor our scaling analysis uncovers
remarkable consistency with 3D-xy critical behavior, revealing that
close to criticality the order parameter is a single complex scalar
as in $^{4}$He. This opens up the window onto the exploration of
the magnetic field induced finite size effect, whereupon the
correlation length transverse to the applied magnetic field $H_{i}$
applied along the $i$-axis cannot grow beyond the limiting magnetic
length $L_{H_{i}}=\left( \Phi _{0}/\left( aH_{i}\right) \right)
^{1/2}$ with $a\simeq 3.12$, related to the average distance between
vortex lines. We find unambiguous evidence for this finite size
effect. It implies that in type II superconductors, such as
MgB$_{2}$, there is the 3D to 1D crossover line $H_{pi}\left(
T\right) =\left( \Phi _{0}/\left( a\xi _{j0}^{-}\xi _{k0}^{-}\right)
\right) (1-T/T_{c})^{4/3}$ with $i\neq j\neq k$ and $\xi
_{i0,j0,k0}^{\pm }$ denotes the critical amplitudes of the
correlation lengths above ($+$) and below ($-$) $T_{c}$ along the
respective axis. Consequently, above $H_{pi}\left( T\right) $ and
$T<T_{c}$ superconductivity is confined to cylinders with diameter
$L_{H_{i}}$(1D). In contrast, above $T_{c}$ and $H_{pi}\left(
T\right) =\left( \Phi _{0}/\left( a\xi _{j0}^{+}\xi _{k0}^{+}\right)
\right) (T/T_{c}-1)^{4/3}$ the uncondensed pairs are confined to
cylinders. Accordingly, there is no continuous phase transition in
the $(H,T)$ -plane along the $H_{c2}$-lines as predicted by the
mean-field treatment.

\end{abstract}

\maketitle

\section{\label{sec:level1}Introduction}
 Since the discovery of superconductivity in
MgB$_{2}$\cite{nagamatsu} many important properties have already
been measured, particularly outside the regime where thermal
fluctuations dominate. The observation of thermal fluctuation
effects have been limited in conventional low-$T_{c}$
superconductors because the large correlation volume makes these
effects very small compared to the mean-field behavior. By contrast,
the high transition temperature $T_{c}$ and small correlation volume
in a variety of cuprate superconductors lead to significant
fluctuation effects\cite{book,parks}. In MgB$_{2}$ the correlation
volume and $T_{c}$ lie between these extremes, suggesting that
fluctuation effects will be observable. Indeed, excess
magnetoconductance\cite{kang}, fluctuation effects in the specific
heat\cite{salamon}, and fluctuating diamagnetic
magnetization\cite{rigamonti} was observed recently in powder
samples. Here we report and analyze reversible magnetization data of
a high quality MgB$_{2}$ single crystal in the vicinity of the zero
field transition temperature, $T_{c}\simeq 38.83$ K, at several
magnetic fields up to $300$ Oe, applied along the $c$-axis. Though
MgB$_{2}$ is a two gap superconductor our scaling analysis uncovers
below $T_{c}$ remarkable consistency with 3D-xy critical behavior,
revealing that the order parameter is a single complex scalar as in
$^{4}$He. The high quality of the single crystal made it possible to
enter this regime. For this reason the magnetic field induced finite
size effect, whereupon the correlation length transverse to the
applied magnetic field cannot grow beyond the limiting magnetic
length $L_{H_{i}}=\left( \Phi _{0}/\left( aH_{i}\right) \right)
^{1/2}$, with the magnetic field $H_{i}$ applied along the $i$-axis
and $a\simeq 3.12$, could be verified and studied in detail.
$L_{H_{i}}$ is related to the average distance between vortex lines.
Indeed, as the magnetic field increases, the density of vortex lines
becomes greater, but this cannot continue indefinitely, the limit is
roughly set on the proximity of vortex lines by the overlapping of
their cores. This finite size effect implies that in type II
superconductors, superconductivity in a magnetic field is confined
to cylinders with diameter $L_{H_{i}}$. Accordingly, there is below
$T_{c}$ the 3D to 1D crossover line $H_{pi}\left( T\right)
=\left( \Phi _{0}/\left( a\xi _{j0}^{-}\xi _{k0}^{-}\right) \right)
(1-T/T_{c})^{4/3}$ with $i\neq j\neq k$. $\xi _{i0,j0,k0}^{\pm }$
denotes the critical amplitudes of the correlation lengths above
($+$) and below ($-$) $T_{c}$ along the respective axis. It
circumvents the occurrence of the continuous phase transition in the
$(H,T)$ -plane along the $H_{c2}$-lines predicted by the mean-field
treatment. Furthermore, our analysis of the magnetization data of
Lascialfari \textit{et al}.\cite{rigamonti} taken on a MgB$_{2}$
powder sample also confirms that there is a magnetic field induced
finite size effect above $T_{c}$ as well. It leads to the line
$H_{pi}\left( T\right) =\left( \Phi _{0}/\left( a\xi _{j0}^{+}\xi
_{k0}^{+}\right) \right) (T/T_{c}-1)^{4/3}$, where the 3D to 1D
crossover occurs and the uncondensed pairs are forced to confine in
cylinders.

The paper is organized as follows: Next we sketch the scaling theory
appropriate for a neutral type II superconductor with a single
complex scalar order parameter falling in the absence of a magnetic
field onto the 3D-xy universality class. Section II is devoted to
experimental details, the presentation of our magnetization data for
$T\lesssim T_{c}$, their analysis by means of the scaling theory and
the analysis of the magnetization data of Lascialfari \textit{et
al}.\cite{rigamonti} taken on a MgB$_{2}$ powder sample for
$T\gtrsim T_{c}$.

Though MgB$_{2}$ is a two gap superconductor an effective one gap
description appears to apply sufficiently close to
$T_{c}$\cite{dao}. As we concentrate on the effects of thermal
fluctuations in the presence of comparatively low magnetic fields we
adopt this effective one gap description. Accordingly, the order
parameter is assumed to be a single complex scalar. To derive the
scaling form of the magnetization in the fluctuation dominated
regime we note that the scaling of the magnetic field is in terms of
the number of flux quanta per correlation area. Thus, when the
thermal fluctuations of the order parameter dominate the singular
part of the free energy per unit volume of a homogeneous system
scales as \cite{book,parks,tsjh2,ffh,tsda,tshkws,tseuro,tsjh}
\begin{equation}
f_{s}=\frac{Q^{\pm }k_{B}T}{\xi _{ab}^{2}\xi _{c}}G^{\pm }\left(z\right) =\frac{Q^{\pm }k_{B}T\gamma }{\xi _{ab}^{3}}G^{\pm }\left(z\right), ~z=\frac{H_{c}\xi _{ab}^{2}}{\Phi _{0}}.
\label{eq1}
\end{equation}
$Q^{\pm }$ is a universal constant and $G^{\pm }\left( z\right) $ a
universal scaling function of its argument, with $G^{\pm }\left(
z=0\right) =1$. $\gamma =\xi _{ab}/\xi _{c}$ denotes the anisotropy,
$\xi _{ab}$ the zero-field in-plane correlation length and $H_{c}$
the magnetic field applied along the $c$-axis. Approaching $T_{c}$
the in-plane correlation length diverges as
\begin{equation}
\xi _{ab}=\xi _{ab0}^{\pm }\left\vert t\right\vert ^{-\nu },~
t=T/T_{c}-1,~\pm =sgn(t).  \label{eq2}
\end{equation}%
Supposing that 3D-xy fluctuations dominate the critical exponents
are given by \cite{pelissetto}
\begin{equation}
\nu \simeq 0.671\simeq 2/3,~\alpha =2\nu -3\simeq -0.013,
\label{eq3}
\end{equation}%
and there are the universal critical amplitude relations \cite%
{book,parks,ffh,tsda,tshkws, pelissetto}%
\begin{equation}
\frac{\xi _{ab0}^{-}}{\xi _{ab0}^{+}}=\frac{\xi _{c0}^{-}}{\xi
_{c0}^{+}}\simeq 2.21,~\frac{Q^{-}}{Q^{+}}\simeq 11.5,~\frac{A^{+}}{A^{-}}=1.07,  \label{eq4}
\end{equation}
and
\begin{eqnarray}
A^{-}\xi _{a0}^{-}\xi _{b0}^{-}\xi _{c0}^{-} &\simeq &A^{-}\left(\xi _{ab0}^{-}\right) ^{2}\xi _{c0}^{-}=\frac{A^{-}\left( \xi_{ab0}^{-}\right)
^{3}}{\gamma }=\left( R^{-}\right) ^{3} \nonumber \\
R^{-}\simeq 0.815,  \label{eq5}
\end{eqnarray}
where $A^{\pm }$ is the critical amplitude of the specific heat
singularity,
defined as%
\begin{equation}
c=\left( A^{\pm }/\alpha \right) \left\vert t\right\vert ^{-\alpha
}+B. \label{eq5a}
\end{equation}
Furthermore, in the 3D-xy universality class $T_{c}$, $\xi
_{c0}^{-}$ and the critical amplitude of the in-plane penetration
depth $\lambda _{ab0}$
are not independent but related by the universal relation \cite%
{book,parks,ffh,tsda,tshkws,pelissetto},
\begin{equation}
k_{B}T_{c}=\frac{\Phi _{0}^{2}}{16\pi ^{3}}\frac{\xi
_{c0}^{-}}{\lambda _{ab0}^{2}}=\frac{\Phi _{0}^{2}}{16\pi
^{3}}\frac{\xi _{ab0}^{-}}{\gamma \lambda _{ab0}^{2}}.  \label{eq6}
\end{equation}
From the singular part of the free energy per unit volume given by (\ref{eq1}) we derive for the magnetization per unit volume
$m=M/V=-\partial f_{s}/\partial H$ the scaling form
\[\frac{m}{TH_c^{1/2}}=-\frac{Q^{\pm }k_{B}\xi _{ab}}{\Phi _{0}^{3/2}\xi
_{c}} F^{\pm }\left( z\right) ,~F^{\pm }\left( z\right)
=z^{-1/2}\frac{ dG^{\pm }}{dz},\]
\begin{equation}
z=x^{-1/2\nu }=\frac{\left( \xi _{ab0}^{\pm }\right) ^{2}\left\vert
t\right\vert ^{-2\nu }H_{c}}{\Phi _{0}}.  \label{eq7}
\end{equation}
In terms of the variable $x$ this scaling form is similar to
Prange's \cite{prange} result for Gaussian fluctuations. More
generally, the existence of the magnetization at $T_{c}$, of the
penetration depth below $T_{c}$ and of the magnetic susceptibility
above $T_{c}$ imply the following asymptotic forms of the scaling
function \cite{book,parks,tsjh2,tseuro,tsjh}
\begin{eqnarray}
Q^{\pm }\left. \frac{1}{\sqrt{z}}\frac{dG^{\pm }}{dz}\right\vert
_{z\rightarrow \infty } &=&Q^{\pm }c_{\infty }^{\pm },  \nonumber \\
Q^{-}\left. \frac{dG^{-}}{dz}\right\vert _{z\rightarrow 0}
&=&Q^{-}c_{0}^{-}\left( \ln \left( z\right) +c_{1}\right) ,  \nonumber \\
Q^{+}\left. \frac{1}{z}\frac{dG^{+}}{dz}\right\vert _{z\rightarrow
0} &=&Q^{+}c_{0}^{+},  \label{eq8}
\end{eqnarray}
with the universal coefficients \cite{book,tsjh2}
\begin{equation}
Q^{-}c_{0}^{-}\simeq -0.7,~Q^{+}c_{0}^{+}\simeq 0.9,~q=Q^{\pm }c_{\infty }^{\pm }\simeq 0.5.  \label{eq9}
\end{equation}
The scaling form (\ref{eq7}) with the limits (\ref{eq8}), together
with the critical exponents (\ref{eq3}) and the universal
relations (\ref{eq4}) and (\ref{eq6}) are characteristic critical
properties of an extreme type II\ superconductor. They provide the
basis to extract from experimental data the doping dependence of the
non-universal critical properties, including the transition
temperature $T_{c}$, the critical amplitudes of correlation lengths
$\xi _{ab0,c0}^{\pm }$, the anisotropy $\gamma $, \textit{etc}.,
while the universal relations are independent of the doping level.

In practice, however, there are limitations set by the presence of
disorder, inhomogeneities and the magnetic field induced finite size
effect. Nevertheless, as cuprate superconductors are concerned there
is considerable evidence for 3D-xy critical behavior, except for a
rounded transition close to $T_{c}$
\cite{book,parks,tsda,tshkws,tseuro,tsjh,hub,babic,ohl,kamal,jacc,kamal2,pasler,roulin,ts07}.
As disorder is concerned there is the Harris criterion
\cite{harris}, which states that short-range correlated and
uncorrelated disorder is irrelevant at the unperturbed critical
point, provided that the specific heat exponent $\alpha $ is
negative. Since in the 3D-xy universality class $\alpha $ is
negative (\ref{eq3}), disorder is not expected to play an
essential role. However, when superconductivity is restricted to
homogeneous domains of finite spatial extent $L_{ab,c}$, the system
is inhomogeneous and the resulting rounded transition uncovers a
finite size effect \cite{cardy,privman} because the correlation
lengths $\xi _{ab,c}=\xi _{ab0,c0}^{\pm }\left\vert t\right\vert
^{-\nu }$ cannot grow beyond $L_{ab,c}$, the respective extent of
the homogenous domains. Hence, as long as $\xi _{ab,c}<L_{ab,c}$ the
critical properties of the fictitious homogeneous system can be
explored. There is considerable evidence that this scenario accounts
for the rounded transition seen in the specific heat \cite{book} and
the magnetic penetration depths \cite{tsdan}. In type II
superconductors, exposed to a magnetic field $H_{i}$, there is an
additional limiting length scale $L_{H_{i}}=\sqrt{\Phi _{0}/\left(
aH_{i}\right) }$ with $a\simeq 3.12$\cite{bled}, related to the
average distance between vortex
lines\cite{parks,bled,haussmann,lortz}. Indeed, as the density of
vortex lines becomes greater with increasing magnetic field, this
cannot continue indefinitely. The limit is roughly set on the
proximity of vortex lines by the overlapping of their cores. Due to
these limiting lengths the correlation lengths cannot grow
beyond\cite{bled}
\begin{eqnarray}
\xi _{i}\left( t_{p}\right) =\xi _{0i}^{\pm }\left\vert
t_{p}\right\vert
^{-\nu }=L_{i},   \nonumber\\
\sqrt{\xi _{i}\left( t_{p}\right) \xi _{j}\left( t_{p}\right) }
=\sqrt{\xi _{0i}^{\pm }\xi _{0j}^{\pm }}\left\vert
t_{p}\right\vert ^{-\nu } =\sqrt{\Phi _{0}/\left( aH_{k}\right) }=L_{H_{k}},  \label{eq10}
\end{eqnarray}
where $i\neq j\neq k$. As the magnetization is concerned the
inhomogeneity induced finite size effect is expected to set in close
to $T_{c}$ where $\xi _{ab,c}$ approaches $L_{ab,c}$, while for a
field applied along the c-axis, the magnetic finite size effect
dominates when $L_{H_{c}}=\sqrt{\Phi _{0}/\left( aH_{c}\right)
}\lesssim L_{ab}$. Accordingly, sufficiently extended magnetization
measurements are not expected to provide estimates for the critical
properties of the associated fictitious homogeneous system only, but
do have the potential to uncover inhomogeneities giving rise to a
finite size effect as well. As a unique size of the homogeneous
domains is unlikely, the smallest extent will set the scale where
the growth of the respective correlation length starts to deviate
from the critical behavior of the homogenous counterpart.

To recognize the implications of the magnetic field induced finite
size effect, it is instructive to note that the scaling form of the
singular part of the free energy per unit volume, (\ref{eq1}),
is formally equivalent to an uncharged superfluid, such as $^{4}$He,
constrained to a cylinder of diameter $L_{H_{c}}=\left( \Phi
_{0}/(aH_{c})\right) ^{1/2}$. Indeed, the finite size scaling theory
predicts, that in a system confined to a barlike geometry, $L\cdot L\cdot
H$, with $H\rightarrow \infty $, an observable $ O(t,L) $ scales
as\cite{cardy,privman,nho}
\begin{equation}
\frac{O\left( t,L\right) }{O\left( t,\infty \right) }=f_{O}\left(
y\right) ,~y=\xi \left( t\right) /L,  \label{eq11}
\end{equation}
where $f(y)$ is the finite size scaling function. As in the confined
system a 3D to 1D crossover occurs, there is a rounded transition
only. Indeed, because the correlation length $\xi \left( t\right) $
cannot grow beyond $L$ there is a rounded transition at
\begin{eqnarray}
T_{p} &=&T_{c}\left( 1-\left( \frac{\xi _{0}^{-}}{L}\right) ^{1/\nu
}\right)
:T<T_{c}.  \nonumber \\
T_{p} &=&T_{c}\left( 1+\left( \frac{\xi _{0}^{+}}{L}\right) ^{1/\nu
}\right) :T>T_{c}.  \label{eq12}
\end{eqnarray}
The resulting rounding of the specific heat singularity and the
shift of the smeared peak from $T_{c}$ to $T_{p}$ is well confirmed
in $^{4}$He\cite{coleman,gasparini}. In superconductors the specific
heat adopts with (\ref{eq5a}) and (\ref{eq11}) the finite size
scaling form
\begin{equation}
c\left( t,L_{H_{c}}\right) =\frac{A^{-}}{\alpha }\left\vert
t\right\vert ^{-\alpha }f_{c}\left( tL_{H_{c}}^{1/\nu }\right)
,~\nu \simeq2/3, \label{eq13}
\end{equation}
where%
\begin{equation}
f_{c}\left( tL_{H_{c}}^{1/\nu }\right) =\left\{
\begin{array}{c}
1~~~~~~~~~~~~~~~~: tL_{H_{c}}^{1/\nu
}=0~~~: t\leq 0 \\
c_{\infty }^{-}\left( tL_{H_{c}}^{1/\nu }\right) ^{\alpha }~:
tL_{H_{c}}^{1/\nu }\rightarrow \infty ~:  t<0%
\end{array}%
\right.  \label{eq14}
\end{equation}
Invoking (\ref{eq12}) in the form $\left\vert t_{p}\right\vert
=\left( \xi _{ab0}^{-}/L_{Hc}\right) ^{1/2\nu }$, the height of the
rounded specific heat peak at $T_{p}$ vanishes then as
\begin{eqnarray}
c\left( T_{p}\right)  &=&\frac{A^{-}}{\alpha }\left\vert
t_{p}\right\vert ^{-\alpha }f_{c}\left( \left( \xi _{ab0}^{-}\right)
^{1/\nu }\right)
\\
&=&\frac{A^{-}}{\alpha }\left( \frac{\left( \xi _{ab0}^{-}\right)
^{2}a}{\Phi _{0}}\right) ^{-\alpha /2\nu }f_{c}\left( \left( \xi
_{ab0}^{-}\right) ^{1/\nu }\right) H_{c}^{-\alpha /2\nu }\nonumber,
\label{eq15}
\end{eqnarray}
because $\alpha <0$ (\ref{eq3}). The resulting shift and
reduction of the rounded specific heat peak with increasing magnetic
field is in a variety of type II superconductors\cite{bled},
including MgB$_{2}$\cite{lyard,salamon}, qualitatively well confirmed.\\
Furthermore, (\ref{eq13}) yields with Maxwell's relation
\begin{equation}
\frac{\partial(C/T)}{\partial H_c}\bigg|_T=\frac{\partial^2M}{\partial T^2}\bigg|_{H_c} \label{maxwell1}
\end{equation}
the scaling form
\begin{equation}
\frac{\partial(c/T)}{\partial H_c}=\frac{\partial^2m}{\partial T^2}=-\frac{k_BA^{\pm}}{2\alpha \nu T}H_c^{-1-\alpha /2\nu}|x|^{1-\alpha}\frac{\partial f_c^{\pm}}{\partial x}. \label{maxwell2}
\end{equation}

\section{\label{sec:level1}Experiment, results and analysis}

The nearly rectangular shaped MgB$_{2}$ single crystal investigated
here was fabricated by high-pressure synthesis described in detail
elsewhere\cite{karpinski}. Its calculated volume is $1.4\cdot
10^{-5}$ cm$^{3}$ and agrees with susceptibility measurements in the
Meissner state with the calculated shape factor $0.81$. The magnetic
moment was measured by a commercial Quantum Design DC-SQUID
magnetometer MPMS XL allowing to achieve a temperature resolution up
to $0.01$ K. The installed reciprocating sample option (RSO) allows to measure magnetic
moments down to $10^{-8}$ emu. In our sample this allows to detect
the magnetic moment near $T_{c}$ down to $25$ Oe. The applied
magnetic field was oriented along the $c$-axis of the sample. After
applying the magnetic field well below $T_{c}$ it was kept constant
and the magnetic moment of the sample was measured at a stabilized
temperature by moving the sample with a frequency of $0.5$ Hz
through a set of detection coils. The diamagnetic magnetization,
$M=mV$, was then obtained by subtracting $M_{b}=5\cdot 10^{-8}H$
emu, the temperature independent paramagnetic and sample holder
contributions. Zero-field cooled (ZFC) magnetization curves have
been compared to field cooled (FC) data, obtained by cooling to a
given temperature in the presence of different fields. Here we
concentrate on the reversible regime (see figure 1) close to $T_{c}$. Due to the
small volume of the sample its magnetic moment can be reliably
detected only below and slightly above $T_{c}$. For this reason we
concentrate on the fluctuation effects below and at $T_{c}$.
\begin{figure}[t!]
\centering
\vspace{-0.6cm}
\includegraphics[width=0.7\linewidth]{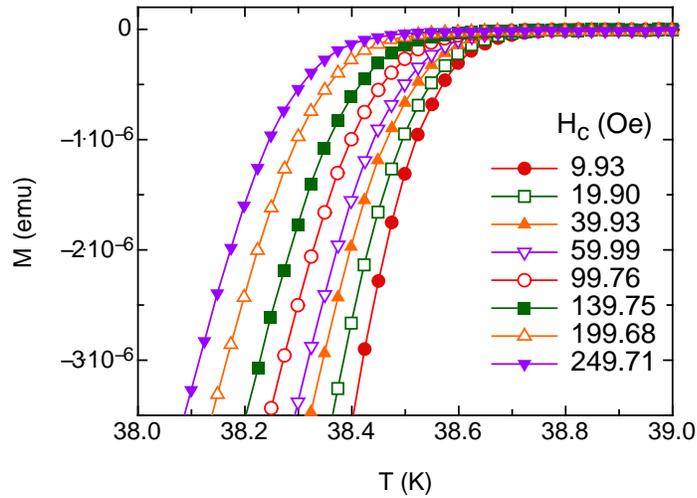}
\vspace{0cm} \caption{Measured magnetic moment of the studied MgB$_2$ single crystals for different magnetic fields applied along the crystals $c$-axis. The lines are guides to the eye. For clarity not all measured fields are shown.}
\label{fig1}
\end{figure}

To estimate $T_{c}$ from the magnetization data $m\left(
T,H_{c}\right) $ we invoke the limit $z\rightarrow \infty $. Here
the scaling form (\ref{eq7}) reduces with (\ref{eq8}) and
(\ref{eq9}) to
\begin{equation}
\frac{m}{H_{c}^{1/2}}=-\frac{k_{B}q}{\Phi _{0}^{3/2}}\frac{\xi
_{ab}}{\xi _{c}}T,~q=Q^{\pm }c_{\infty }^{\pm }\simeq 0.5.
\label{eq16}
\end{equation}
$Q^{+}c_{\infty }^{+}=Q^{-}c_{\infty }^{-}$ follows from the fact
that $ m/H_{c}^{1/2}$ adopts at the zero-field transition
temperature $T_{c}$ a unique value. Here the curves $m/H_{c}^{1/2}$
vs. $T$ taken at different fields $H_{c}$ should cross and
$m/H_{c}^{1/2}\gamma T_{c}$ adopts the universal value
\begin{equation}
\frac{m\xi _{c}\left( T_{c}\right) }{H_{c}^{1/2}T_{c}\xi _{ab}\left(
T_{c}\right) }=-\frac{k_{B}q}{\Phi _{0}^{3/2}}.  \label{eq17}
\end{equation}
Accordingly, the location of a crossing point in $m/H_{c}^{1/2}$
vs. $T$ provides an estimate for the 3D transition
temperature and the factor of proportionality in $m/T_{c}$ vs. $H_{c}^{1/2} $ probes the
anisotropy $\gamma =\xi _{ab}\left( T_{c}\right) /\xi _{c}\left(
T_{c}\right) $. From Fig. \ref{fig1} showing $m/H_{c}^{1/2}$ vs. $T$
we derive the estimate $T_{c}\simeq 38.83$ K and (\ref{eq17})
yields with $m/(T_{c}H_{c}^{1/2})\approx 1.4\cdot 10^{-6}$(emu
cm$^{-3}$K$^{-1}$Oe$^{-1/2} $) for the anisotropy the value
\begin{equation}
\xi _{ab}\left( T_{c}\right) /\xi _{c}\left( T_{c}\right) \approx
1.9. \label{eq18}
\end{equation}
In a homogeneous system where the correlation lengths diverge at
$T_{c}$ as $\xi _{ab,c}=\xi _{ab0,c0}^{\pm }\left\vert t\right\vert
^{-\nu }$ with $\nu \simeq 2/3$, whereupon $\xi _{ab}\left(
T_{c}\right) /\xi _{c}\left( T_{c}\right) $ corresponds to the anisotropy $\gamma =\xi _{ab0}^{\pm }/\xi
_{c0}^{\pm }$. In contrast, in an inhomogeneous system, consisting
of homogenous domains of spatial extent $L_{ab,c}$ this ratio probes
$\xi _{ab}\left( T_{c}\right) /\xi _{c}\left( T_{c}\right)
=L_{ab}/L_{c}$, because the correlation lengths cannot exceed the
homogenous domains. Nevertheless $\xi _{ab}\left( T_{c}\right) /\xi
_{c}\left( T_{c}\right) \approx 1.9$ is close to $\gamma \simeq 2$,
the estimate obtained near $T_{c}$ with torque
magnetometry\cite{angst}.
\begin{figure}[t!]
\centering
\vspace{-0.6cm}
\includegraphics[width=0.7\linewidth]{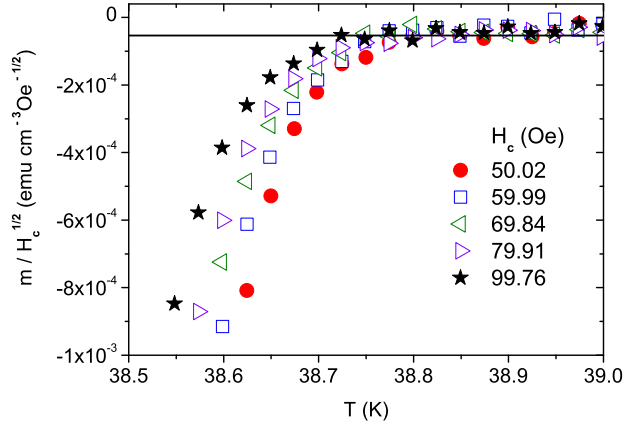}
\vspace{-0.6cm} \caption{$m/H_{c}^{1/2}$ vs. $T$ for a MgB$_{2}$
single crystal with the magnetic field $H_{c}$ applied along the
$c$-axis. The solid line is $m/(T_{c}H_{c}^{1/2})\approx -1.4\cdot
10^{-6}$(emu cm$^{-3}$K$^{-1}$Oe$^{-1/2}$) with $T_{c}=38.83$ K.}
\label{fig1}
\end{figure}

\begin{figure}[b!]
\centering
\vspace{-0.6cm}
\includegraphics[width=0.7\linewidth]{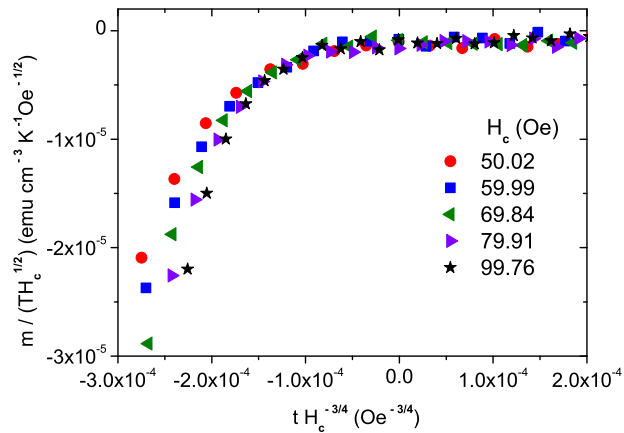}
\vspace{-0.60cm} \caption{Scaling plot $m/(TH^{1/2})$ vs.
$tH_{c}^{-3/4}$.} \label{fig2}
\end{figure}

According to the scaling form (\ref{eq7}) consistency with critical
behavior also requires that for low fields the data plotted as
$m/(TH_c^{1/2})$ vs. $tH_{c}^{-3/4}$ should collapse near
$tH_{c}^{-3/4}\rightarrow 0$ on a single curve. Evidence for this
collapse emerges from Fig. \ref{fig2}.

\begin{figure}[t!]
\centering
\vspace{-0.6cm}
\includegraphics[width=0.7\linewidth]{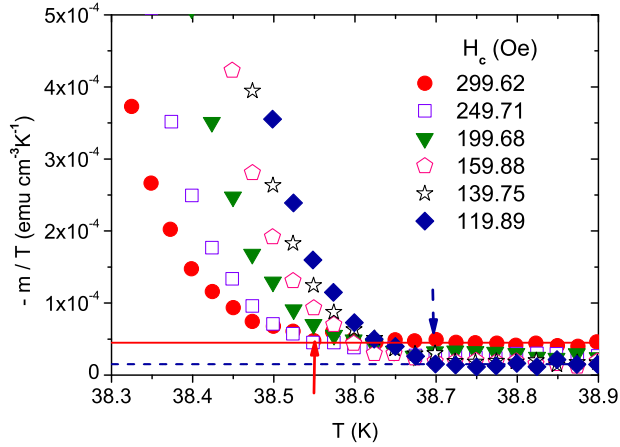}
\vspace{-0.60cm} \caption{$-m/T$ vs. $T$ for various applied
magnetic fields. The solid line indicates
$-m/T_{p}=4.5\cdot10^{-5}$(emu cm$^{-3}$K$^{-1}$) at $H_{c}=299.2$
Oe, where $T_{p}\simeq 38.55$ K and the dashed one
$-m/T_{p}=1.5\cdot10^{-5}$ (emu cm$^{-3}$K$^{-1}$) at $H_{c}=119.9$
Oe, where $T_{p}\simeq 38.7 $ K. The arrows mark the respective
$T_{p}$'s.} \label{fig3}
\end{figure}

Because the limiting magnetic length, $L_{H_{c}}=\sqrt{\Phi
_{0}/\left( aH_{c}\right) }$, decreases with increasing field this
scaling behavior does no longer apply at higher fields. Indeed, with
increasing field $L_{H_{c}}=\sqrt{\Phi _{0}/\left( aH_{c}\right) }$
approaches $\xi _{ab}$ and when $\xi _{ab}\left( T_{p}\right)
=L_{H_{c}}$ the scaling form (\ref{eq16}) reduces to
\begin{equation}
\frac{m}{T_{p}}\simeq -0.5\frac{k_{B}}{\Phi _{0}^{3/2}}\frac{\xi
_{ab}\left( T_{p}\right) }{\xi _{c}\left( T_{p}\right)
}H_{c}^{1/2}=-0.5\frac{k_{B}}{\Phi _{0}a^{1/2}}\frac{1}{\xi
_{c}\left( T_{p}\right) }, \label{eq19}
\end{equation}
where
\begin{equation}
T_{p}=T_{c}\left( 1-\left( \frac{aH_{c}\left( \xi
_{ab0}^{-}\right) ^{2}}{\Phi _{0}}\right) ^{3/4}\right)=T_{c}\left( 1-\left( \frac{\xi _{ab0}^{-}}{L_{H_{c}}}\right)
^{3/2}\right) ,  \label{eq20}
\end{equation}
in analogy to (\ref{eq12}), the expression for $^{4}$He
constrained below $T_{c}$ to cylinders of diameter $L$. Accordingly,
in sufficiently high fields the magnetic field induced finite size
effect is predicted to eliminate the characteristic critical field
dependence, $-m/T_{c}\propto H_{c}^{1/2}$, emerging from Fig.
\ref{fig1}, because the in-plane correlation length $\xi _{ab}$
cannot grow beyond $L_{H_{c}}$. A glance to Fig. \ref{fig3}, showing
$-m/T$ vs. $T$ for various applied magnetic fields in the
range from $120$ to $300$ Oe reveals that this prediction is well
confirmed in this field range. Indeed, $-m/T$ levels off above
$T=T_{p}$ and the magnitude of $-m/T_{p}$ is controlled by $\xi
_{c}\left( T_{p}\right)$.

\begin{figure}[t!]
\centering
\vspace{-0.6cm}
\includegraphics[width=0.7\linewidth]{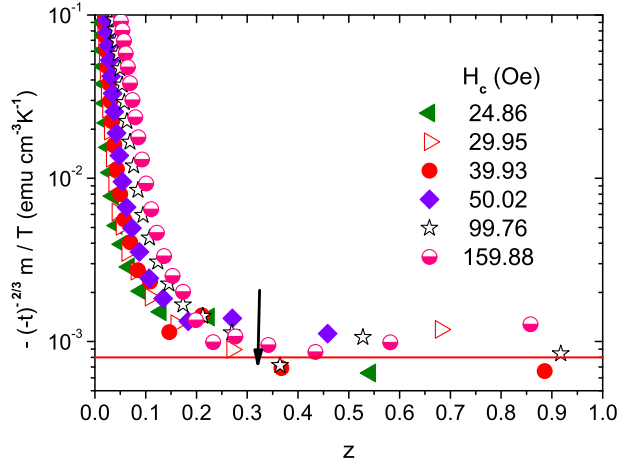}
\vspace{-0.6cm} \caption{$\left\vert t\right\vert ^{-2/3}m/T$
vs. $z$ for various fields. The solid line is $\left\vert
t\right\vert ^{-2/3}m/T=8\cdot 10^{-4}$(emu cm$^{-3}$K$^{-1}$) and
the arrow marks $z=1/a\simeq 0.32$.} \label{fig4}
\end{figure}

Using Eq.(\ref{eq20}), $T_{p}\left( H_{c}=299.2\textrm{ Oe}\right)
\simeq 38.55$ K and $T_{p}\left( H_{c}=119.9 \textrm{ Oe}\right)
\simeq 38.7$ K we obtain for the critical amplitude of the in-plane
correlation length the estimate
\begin{equation}
\xi _{ab0}^{-}\simeq 52\textrm{ \AA .}  \label{eq21}
\end{equation}
On this basis the dependence of $m/(TH_{c}^{1/2})$ on the scaling
variable $z=\left( \xi _{ab0}^{-}\right) ^{2}\left\vert t\right\vert
^{-4/3}H_{c}/\Phi _{0}$ is then readily calculated. When the
magnetic field induced finite size effect scenario holds true, the
effective range of the scaling variable is restricted to
\begin{equation}
z\leq 1/a\simeq 0.32,  \label{eq22}
\end{equation}
because the correlation length cannot exceed $\xi _{ab}=L_{H_{c}}$.
As a consequence (\ref{eq7}) reduces for $z\gtrsim 0.32$ to
\begin{equation}
\left\vert t\right\vert ^{-2/3}\frac{m}{T}=-\frac{k_{B}}{\Phi
_{0}\xi _{c0}^{-}}Q^{-}\left. \frac{dG^{-}}{dz}\right\vert _{z=1/a}.
\label{eq23}
\end{equation}
Accordingly, in the plot $\left\vert t\right\vert ^{-2/3}m/T$
\textit{vs}. $z $ the data should collapse and level off for
$z\gtrsim 0.32$. From Fig. \ref{fig4}, showing this scaling plot, it
is seen that this behavior is well confirmed down to $H_{c}=24.86$
Oe, whereupon we obtain for $L_{ab}$, the spatial extent of the
homogenous domains in the $ab$-plane, the lower bound
\begin{equation}
L_{ab}=\left( \frac{\Phi _{0}}{aH_{c}}\right) ^{1/2}\geq
5.2\cdot10^{-5} \textrm{ cm,}  \label{eq24}
\end{equation}
revealing the high quality of the sample.

\begin{figure}[t!]
\centering
\vspace{-0.6cm}
\includegraphics[width=0.7\linewidth]{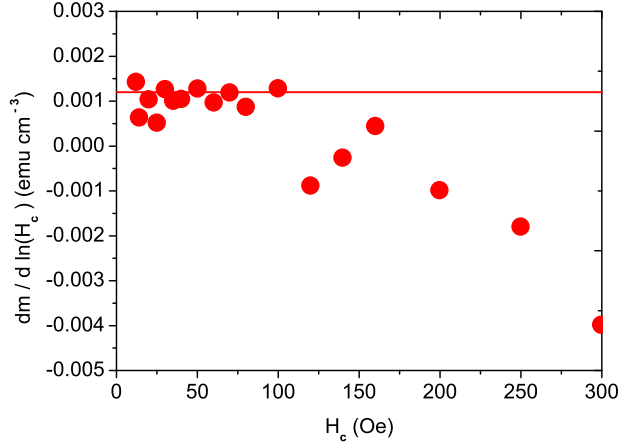}
\vspace{-0.6cm} \caption{$dm/d\ln \left( H_{c}\right) $ vs.
$H_{c}$ at $T=38.7$ K. The solid line is $dm/d\ln \left(
H_{c}\right) =1.2\cdot 10^{-3}$(emu cm$^{-3}\ln $(Oe)$^{-1}$).}
\label{fig5}
\end{figure}

To check the estimates for the critical amplitudes of the
correlation lengths, we invoke, using (\ref{eq7}), (\ref{eq8})
and (\ref{eq9}), the limiting behavior
\begin{equation}
\frac{dm}{d\ln \left( H_{c}\right) }=0.7\frac{k_{B}T}{\Phi _{0}\xi
_{c}}, \label{eq25}
\end{equation}
applicable for $z\rightarrow 0$. From the plot $dm/d\ln \left(
H_{c}\right) $ vs. $H_{c}$ at $T=38.7$ K, shown in Fig.
\ref{fig5} and $dm/d\ln \left( H_{c}\right) =1.2\cdot 10^{-3}$ (emu
cm$^{-3}$ ln(Oe)$^{-1}$) we obtain for the critical amplitude of the $c$-axis
correlation length the estimate
\begin{equation}
\xi _{c0}^{-}\simeq 33\textrm{ \AA ,}  \label{eq26}
\end{equation}
in reasonable agreement with $\xi _{abo}^{-}/\gamma \simeq
52$ \AA$/1.9\simeq 27$ \AA . Note that at this temperature and $\xi
_{abo}^{-}\simeq 52$ \AA\ the limit $z\rightarrow 0$ is attained
because $z=2.74\cdot10^{-3}H_{c}$, with $H_{c}$ in Oe. Together with
the universal relation (\ref{eq6}), $\xi _{c0}^{-}\simeq 33$ \AA ,
yields for the critical amplitude of the in-plane penetration depth,
$\lambda _{ab0}$, and the Ginzburg parameter, $\kappa _{ab0}$, the
estimates
\begin{equation}
\lambda _{ab0}\simeq 7.3\cdot \textrm{10}^{-5}\textrm{ cm, }\kappa
_{ab0}=\lambda _{ab0}/\xi _{ab0}^{-}\simeq 140,  \label{eq27}
\end{equation}
which apply very close to $T_{c}$. Unfortunately, the available
magnetic penetration depth data does not enter this
regime\cite{panagopoulos,dicastro}.

To explore the evidence for an inhomogeneity induced finite size
effect, attributable to a system consisting of homogeneous domains
of finite extent, we rewrite the scaling form (\ref{eq7}) with the
aid of (\ref{eq8}) in the form
\begin{equation}
\frac{m}{T}=-\frac{k_{B}}{\Phi _{0}\xi _{c}}Q^{-}\frac{dG^{-}}{dz}=-\left\vert t\right\vert ^{2/3}\frac{k_{B}}{\Phi _{0}\xi
_{c0}}\left. Q^{-}\frac{dG^{-}}{dz}\right\vert_{z=H_{c}L_{ab}^{2}/\Phi _{0}},  \label{eq28}
\end{equation}
because $\xi _{ab}$ cannot grow beyond $L_{ab}$, the extent of the
homogeneous domains in the $ab$-plane. However, sufficiently close
to $T_{c}$, $\xi _{c}$ approaches $L_{c}$, the extent of the
homogeneous domains along the $c$-axis. Here this scaling form
reduces to
\begin{eqnarray}
\frac{m}{T} &=&-f_{0}\left( H_{c}\right) ,  \nonumber \\
\textrm{ }f_{0}\left( H_{c}\right)  &=&\frac{k_{B}}{\Phi
_{0}L_{c}}\left. Q^{-}\frac{dG^{-}}{dz}\right\vert
_{z=H_{c}L_{ab}^{2}/\Phi _{0}}.  \label{eq29}
\end{eqnarray}
In Fig. 7 we depicted $-\left\vert t\right\vert ^{-2/3}m/T$
vs. $-t$. Apparently, this limiting behavior is attained
roughly below $-t=-t_{pL_{c}}=3\cdot 10^{-4}$, where
\begin{equation}
\xi _{c}\left( t\right) =\xi _{c0}^{-}\left\vert
t_{pL_{c}}\right\vert ^{-2/3}=L_{c}.  \label{eq30}
\end{equation}
With $\xi _{c0}^{-}=\xi _{ab0}^{-}/\gamma \simeq 52$ \AA$/1.9$ we
obtain for $L_{c}$, the $c$-axis extent of the homogenous domains,
the estimate
\begin{equation}
L_{c}\approx 6\cdot 10^{-5}\textrm{ cm,}  \label{eq31}
\end{equation}
which is comparable to the lower bound $L_{ab}\geq 5.2\cdot10^{-5}$ cm (\ref{eq24}), revealing again the high quality of the sample.
\begin{figure}[htb]
\centering
\vspace{-0.6cm}
\includegraphics[width=0.7\linewidth]{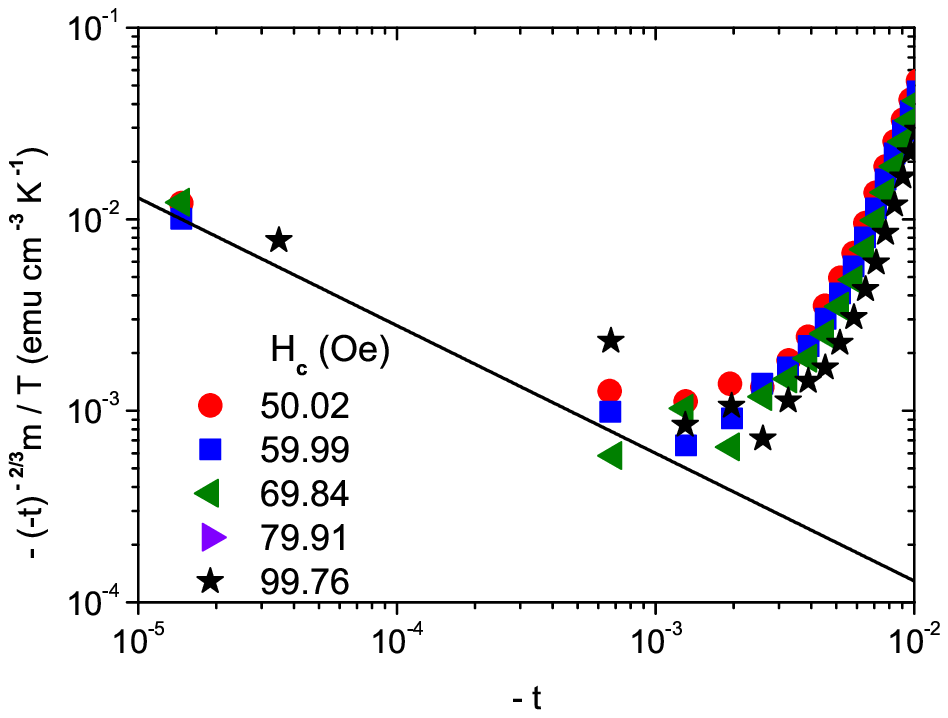}
\vspace{-0.8cm} \caption{$-\left\vert t\right\vert ^{-2/3}m/T$
vs. $-t$. The solid line is $-\left\vert t\right\vert
^{-2/3}m/T=6\cdot $ $10^{-6}\left( -t\right) ^{-2/3}$ (emu
cm$^{-3}$K$^{-1}$).} \label{fig6}
\end{figure}

We have seen that the attainable critical regime is limited by both,
the magnetic field and inhomogeneity induced finite size effects.
The former leads according to (\ref{eq20}) in the $(H,T)$-plane
to the line
\begin{eqnarray}
H_{cp}\left( T\right)  &=&\frac{\Phi _{0}}{a\left( \xi
_{ab0}^{-}\right) ^{2}}\left(1-\frac{T}{T_{c}}\right)^{4/3}:T<T_{c},  \nonumber \\
H_{cp}\left( T\right)  &=&\frac{\Phi _{0}}{a\left( \xi _{ab0}^{-}\right)
^{2}}\left(\frac{T}{T_{c}}-1\right)^{4/3}:T>T_{c},  \label{eq32}
\end{eqnarray}
\begin{figure}[b!]
\centering
\vspace{-0.6cm}
\includegraphics[width=0.7\linewidth]{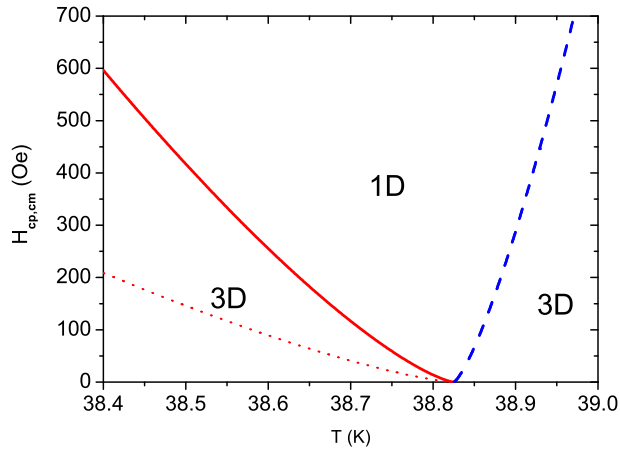}
\vspace{-0.7cm} \caption{Crossover lines $H_{cp}$ and vortex melting line $H_{cm}$ vs. $T$. The 3D to 1D and the 1D to 3D crossover lines $H_{cp}$ follows from (\ref{eq32}) for $\xi _{ab0}^{-}=52$ \AA\ \ (\ref{eq21}), $\xi _{ab0}^{+}=52$ \AA$/2.21\simeq 23.62$ \AA\ (\ref{eq4}) and $T_{c}=38.83$ K. The solid line applies below $T_{c}$ and the dashed line above $T_{c}$. The dotted vortex melting line $H_{cm}$ follows from (\ref{eq37a}) and lies at temperatures below the crossover lines $H_{cp}$.} \label{fig7}
\end{figure}
depicted in Fig. \ref{fig7}. It is a crossover line because for a
fixed temperature, e.g. below $T_{c}$, the limiting length
$L_{H_{c}}=\left( \Phi _{0}/\left( aH_{c}\right) \right) ^{1/2}$
decreases with increasing magnetic field and matches at $H_{cp}$ the
in-plane correlation length $\xi _{ab}$. Here and above $H_{cp}$
superconductivity is then confined to cylinders of diameter
$L_{H_{cp}}$ in the $ab$-plane and height $L_{c}$ along the
$c$-axis. Hence in a homogenous system where $L_{c}=L_{ab}=\infty $
a 3D to 1D crossover takes place. Even in the presence of
inhomogeneities, corresponding to homogeneous domains of extent
$L_{ab,c}$, this holds true when $H_{c}>\Phi _{0}/\left(
aL_{ab}^{2}\right) $ and $-t=1-T/T_{c}>\left( \xi
_{c0}^{-}/L_{c}\right) ^{3/2}$ because the magnetic field induced
finite size effect dominates when $L_{H_{c}}<L_{ab}$ and $\xi
_{c}<L_{c}$. Indeed below $H_{c}=\Phi _{0}/\left( aL_{ab}^{2}\right)
$ and $-t=1-T/T_{c}=\left( \xi _{c0}^{-}/L_{c}\right) ^{3/2}$
superconductivity occurs in finite boxes with extent
$L_{ab}^{2}L_{c}$ and above superconductivity is again confined to
cylinders and their finite height $L_{c}$ is not detected because
$\xi _{c}<L_{c}$. Noting then that in the present case of MgB$_{2}$,
 $L_{ab}\geq
5.2\cdot 10^{-5}$ cm (\ref{eq24}), the 3D to 1D crossover
scenario applies down to fields smaller than $25$ Oe, while the
finite extent of the homogeneous domains along the $c$-axis requires
that $1-T/T_{c}\gtrsim 3\cdot 10^{-4}$ (see Fig. 6), excluding a very
narrow temperature range below $T_{c} $.

\begin{figure}[b!]
\centering
\vspace{-0.6cm}
\includegraphics[width=0.7\linewidth]{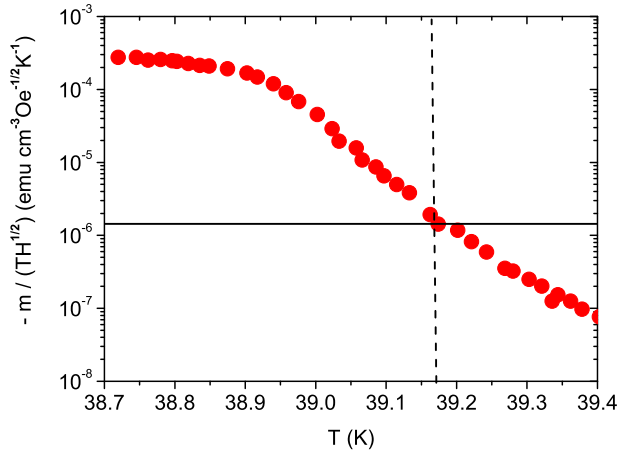}
\vspace{-0.7cm} \caption{$m/\left( TH^{1/2}\right) $ vs.
$T$ at $H=1$ Oe for the MgB$_2$ powder sample of Lascialfari \textit{et
al}.\protect\cite{rigamonti}. The horizontal line is $m/\left(
T_{c}H^{1/2}\right) \simeq -1.44\cdot 10^{-6}$ (emu
cm$^{-3}$K$^{-1}$Oe$^{-1/2}$) and the vertical one marks
$T_{c}\simeq 39.17$ K.} \label{fig8}
\end{figure}

Finally we show that this scenario is also consistent with the
measurements of Lascialfari \textit{et al}.\cite{rigamonti}
performed on powder samples at $T\gtrsim T_{c}$. The rather large
volume of the sample made it possible to explore the critical regime
above $T_{c}$ as well. To demonstrate the consistency with our
analysis we reproduced some data in Fig. \ref{fig8} in terms of
$m/\left( TH^{1/2}\right) $ vs. $T$. For a powder sample we
obtain from (\ref{eq7}), (\ref{eq8}) and (\ref{eq9}) at $T_{c}$
the value
\begin{equation}
\frac{m}{T_{c}H^{1/2}}=-0.5\frac{k_{B}\gamma }{\Phi
_{0}^{3/2}}\left\langle \epsilon \left( \delta \right)
^{3}\right\rangle ,\textrm{ }\gamma =\frac{\xi _{ab}}{\xi _{c}},
\label{eq33}
\end{equation}
where
\begin{equation}
\epsilon \left( \delta \right) =\left( \cos \left( \delta \right)
^{2}+\frac{1}{\gamma ^{2}}\sin \left( \delta \right) ^{2}\right)
^{1/2}. \label{eq34}
\end{equation}
$\delta $ denotes the random orientation of the applied magnetic
field with respect to the $c$-axis and $\left\langle \epsilon \left(
\delta \right) ^{3}\right\rangle $ is the corresponding average. For
$\gamma =1.9$ we obtain $\left\langle \epsilon \left( \delta \right)
^{3}\right\rangle \simeq 0.541$ and with that $m/\left(
T_{c}H^{1/2}\right) \simeq -1.44\cdot 10^{-6}$ (emu
cm$^{-3}$K$^{-1}$Oe$^{-1/2}$). Perfect agreement with our analysis
emerges from Fig. \ref{fig8} for $T_{c}\simeq 39.17$ K, consistent
with the observation of Lascialfari \textit{et al}.\cite{rigamonti}
that in this sample $T_{c}$ is near $39.1$ K. To explore the occurrence of the vortex
melting transition and the 3D to 1D crossover we displayed in Fig. \ref%
{fig10} the data of Lascialfari \textit{et al.}\cite{rigamonti} according to
the scaling form (\ref{maxwell2}). The minimum at $t_{p}H^{-3/4}\simeq
-3.4\cdot 10^{-3}$ Oe$^{-3/4}$ locates the 3D to 1D crossover line, while
the peak at $t_{m}H^{-3/4}\simeq -7.5\cdot 10^{-3}$ Oe$^{-3/4}$ signals the
vortex melting transition. For the ratio of the universal values of the
scaling variable $z$ at the melting and the 1D to 3D crossover line we
obtain the estimate 
\begin{equation}
z_{m}/z_{p}=\left( t_{p}\left( H\right) /t_{m}\left(H\right) \right) ^{4/3}\simeq 0.35, 
\label{eq37a}
\end{equation}
in reasonable agreement with $z_{m}/z_{p}\simeq 0.25$, the value emerging from the specific heat data of Roulin \textit{et al.}\cite{roulin} for YBa$_{2}$Cu$_{3}$O$_{6.97}$. The resulting vortex melting line is included in Fig. \ref{fig7}.

\begin{figure}[htb]
\centering
\vspace{-0.6cm}
\includegraphics[width=0.7\linewidth]{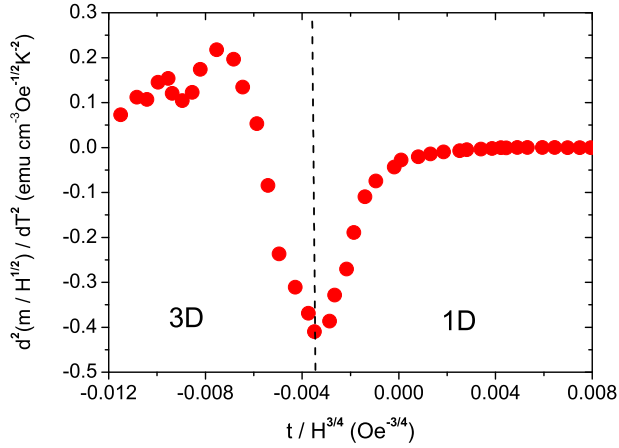}
\vspace{-0.7cm} \caption{$d^{2}\left( m/H^{1/2}\right) /dT^{2}$ \textit{vs}. $t/H^{3/4}$ for
$H=1$ Oe derived from the data of Lascialfari \textit{et al}.\protect\cite
{rigamonti}. The minimum at $t_{p}H^{-3/4}\simeq -3.4\cdot 10^{-3}$ Oe$
^{-3/4}$ locates the 3D to 1D crossover line, while the peak at $%
t_{m}H^{-3/4}\simeq -7.5\cdot 10^{-3}$ Oe$^{-3/4}$ signals the vortex
melting transition.} \label{fig10}
\end{figure}

\begin{figure}[t!]
\centering
\vspace{-0.6cm}
\includegraphics[width=0.7\linewidth]{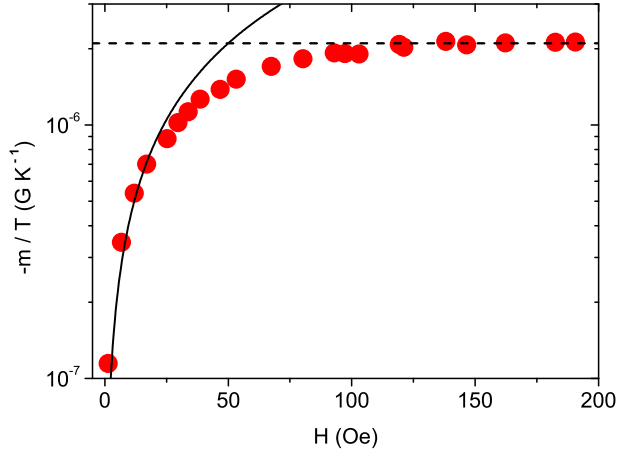}
\vspace{-0.7cm} \caption{$m/T$ vs. $H$ at $T=39.3$ K for the
MgB$_{2}$ powder sample of
Lascialfari \textit{et al}.\protect\cite{rigamonti}. The solid line is $%
m/T=-4.2\cdot 10^{-8}H$ (GK$^{-1}$) and the dashed one
$m/T=-2.1\cdot 10^{-6} $(GK$^{-1}$), marking the saturation due to
the magnetic field induced finite size effect.} \label{fig9}
\end{figure}

At higher fields and fixed temperature, however, a crossover from
$m/T$ $\propto H$ to $m/T=const$ is expected to occur. Indeed,
approaching the limit $z\rightarrow 0$, the scaling form
\begin{equation}
\frac{m}{T}=-0.9\frac{k_{B}\xi _{ab}^{2}}{\Phi _{0}^{2}\xi
_{c}}\left\langle \epsilon \left( \delta \right) ^{2}\right\rangle
H,  \label{eq35}
\end{equation}
applies according to (\ref{eq7}), (\ref{eq8}) and (\ref{eq9}).
As the scaling variable $z$ increases with rising magnetic field it
approaches the value $z=1/a$ where the magnetic field induced finite
size effect sets in. Here the scaling expression (\ref{eq7}) applies
in the form
\begin{equation}
\frac{m}{T}=-\frac{k_{B}}{\Phi _{0}\xi _{c}}\left\langle \epsilon
\left( \delta \right) \right\rangle Q^{+}\left.
\frac{dG^{+}}{dz}\right\vert _{z=1/a,}  \label{eq36}
\end{equation}
for $z\geq 1/a$, where
\begin{equation}
z=\frac{H\xi _{ab}^{2}}{\Phi _{0}}\epsilon \left( \delta \right.)
\label{eq37}
\end{equation}
From Fig. \ref{fig9}, showing $m/T$ vs. $H$ at $T=39.3$ K for the
MgB$_{2}$ powder sample of Lascialfari \textit{et
al}.\cite{rigamonti}, it is seen that this behavior, including the
saturation due to the magnetic field induced finite size effect, is
well confirmed.
Therefore, in analogy to the situation below $T_{c}$, there is a
magnetic field induced finite size effect above $T_{c}$ as well.
However, there is no long range order in this regime so that
uncondensed pairs are forced to confine above $H_{cp}\left( T\right)
$ (see Fig. \ref{fig7}) in cylinders of diameter $L_{H_{cp}}$.

\section{\label{sec:level1}Summary}

To summarize, our scaling analysis of reversible magnetization data
of a MgB$_{2}$ single crystal with the magnetic field applied along
the $c$-axis provided considerable evidence that even in this type
II superconductor the 3D-xy critical regime is experimentally
accessible, provided that the sample is sufficiently homogeneous.
The high quality of our sample allowed to explore the occurrence of
the magnetic field induced finite size effect down to rather low
magnetic fields where 3D-xy fluctuations still dominate. In this
regime we were able to provide rather unambiguous evidence for this
finite size effect. It implies that in type II superconductors, such
as MgB$_{2}$, exposed to a magnetic field superconductivity is
confined to cylinders. Their diameter is given by the limiting
magnetic length $L_{H_{i}}=\left( \Phi _{0}/(aH_{i}\right) )^{1/2}$,
whereupon for a magnetic field applied parallel to the $i$-axis,
there is the line $H_{pi}\left( T\right) =\left( \Phi _{0}/\left(
a\xi _{j0}^{-}\xi _{k0}^{-}\right) \right) (1-T/T_{c})^{4/3}$ with
$i\neq j\neq k$, where below $T_{c}$ a 3D to 1D crossover takes
place. $\xi _{i0,j0,k0}^{-}$ denote the critical amplitudes of the
correlation length below $T_{c}$ along the respective axis.
Accordingly, there is below $T_{c}$ no continuous phase transition
in the $(H,T)$ -plane along the $H_{c2}$-lines as predicted by the
mean-field treatment. Our scaling analysis of the magnetization data
of Lascialfari \textit{et al}.\cite{rigamonti} also confirmed that
the magnetic field induced finite size effect is not restricted to
the superconducting phase ($T<T_{c}$).  Indeed, above $T_{c}$ there
is the line $H_{pi}\left( T\right) =\left( \Phi _{0}/\left( a\xi
_{j0}^{+}\xi _{k0}^{+}\right) \right) (T/T_{c}-1)^{4/3}$ where the
3D to 1D crossover occurs and uncondensed pairs are forced to
confine in cylinders. Furthermore, we have shown that the scaling
analysis of magnetization data also opens a door onto the
ascertainment of the homogeneity of the sample in terms of the
finite size effect arising from the limited extent of the homogenous
domains.

\section{Acknowledgments}

The authors are grateful to J. Roos for the help to prepare the
manuscript and useful comments. This work was supported by the Swiss
National Science Foundation and in part by the NCCR program MaNEP.


\section{References}
\begin{thebibliography}{99}
\bibitem{nagamatsu} Nagamatsu J, Nakagawa N, Maranaka T, Zenitani Y and
Akimitsu J 2001 \textit{Nature (London)} \textbf{410} 63

\bibitem{book} Schneider T and Singer J M 2000 \textit{Phase Transition
Approach To High Temperature Superconductivity} (Imperial College
Press, London)

\bibitem{parks} Schneider T 2004 \textit{The Physics of Superconductors}
edited by K. Bennemann and J. B. Ketterson (Springer, Berlin) p. 111

\bibitem{kang} Kang W N \textit{et al} 2002 \textit{J. Korean Phys. Soc.} \textbf{40}
949

\bibitem{salamon} Park Tuson, Salamon M B, Jung C U, Park Min-Seok,
Kim Kyunghee and Lee Sung-Ik 2002 \textit{Phys. Rev. B} \textbf{66} 134515

\bibitem{rigamonti} Lascialfari A, Mishonov T, Rigamonti A, Tedesco P
and Varlamov A 2002 \textit{Phys. Rev. B} \textbf{65} 180501(R)

\bibitem{dao} Dao V H and Zhitomirski M E 2005 \textit{ Eur. Phys. J. B} \textbf{44}
183

\bibitem{tsjh2} Hofer J, Schneider T, Singer J M, Willemin M, Keller H, Sasagawa T, Kishio K, Conder K and Karpinski J 2000 \textit{Rev. B} \textbf{62} 631

\bibitem{ffh} Fisher D S, Fisher M P A and Huse D A 1991 \textit{Phys. Rev. B}
\textbf{43} 130

\bibitem{tsda} Schneider T and Ariosa D 1992 \textit{Z. Phys. B} \textbf{89} 267

\bibitem{tshkws} Schneider T and Keller H 1993 \textit{Int. J. Mod. Phys. B} \textbf{8} 487

\bibitem{tseuro} Schneider T, Hofer J, Willemin M, Singer J M and Keller H 1998 \textit{Eur. Phys. J. B} \textbf{3} 413

\bibitem{tsjh} Hofer J, Schneider T, Singer J M, Willemin M, Keller H,
Rossel C and Karpinski J 1999 \textit{Pys. Rev. B} \textbf{60} 1332

\bibitem{pelissetto} Pelissetto A and Vicari E 2002 \textit{Physics Reports} \textbf{368} 549

\bibitem{prange} Prange R E 1970 \textit{Phys. Rev. B} \textbf{1} 2349

\bibitem{hub} Hubbard M A, Salamon B and Veal B W 1996 \textit{Physica C} \textbf{259} 309

\bibitem{babic} Bab\'{\i}c D, Cooper J R, Hodby J W and Changkang Chen 1999 \textit{Phys. Rev. B} \textbf{60} 698

\bibitem{ohl} Overend N, Howson M A and Lawrie I D 1994 \textit{Phys. Rev. Lett.} \textbf{72} 3238

\bibitem{kamal} Kamal S, Bonn D A, Goldenfeld N, Hirschfeld P J, Liang R and Hardy W N 1994 \textit{Phys. Rev. Lett.} \textbf{73} 1845

\bibitem{jacc} Jaccard Y, Schneider T, Looquet J P, Williams E J, Martinoli P and Fischer \O 1996 \textit{Europhys. Lett.} \textbf{34} 281

\bibitem{kamal2} Kamal S, Liang R, Hosseini A, Bonn D A and Hardy W N 1998 \textit{Phys. Rev. B} \textbf{58} R8933

\bibitem{pasler} Pasler V, Schweiss P, Meingast Ch, Obst B, W\"{u}hl H, Rykov A I and Tajima S 1998 \textit{Phys. Rev. Lett.} \textbf{81} 1094

\bibitem{roulin} Roulin M, Junod A and Walker E 1998 \textit{Physica C} \textbf{296}
137

\bibitem{ts07} Schneider T 2007 \textit{Phys. Rev. B} \textbf{75} 174517

\bibitem{harris} Harris A B 1974 \textit{J. Phys. C} \textbf{7} 1671

\bibitem{cardy} Cardy J L ed. 1988 \textit{Finite-Size Scaling} North
Holland Amsterdam

\bibitem{privman} Privman V 1990 \textit{Finite Size Scaling and Numerical
Simulations of Statistical Systems, World Scientific} NJ

\bibitem{tsdan} Schneider T and Di Castro D 2004 \textit{Phys. Rev. B} \textbf{69}
024502

\bibitem{bled} Schneider T \textit{Journal of Superconductivity} 2004 \textbf{17} 41

\bibitem{haussmann} Haussmann R 1999 \textit{Phys. Rev. B} \textbf{60} 12373

\bibitem{lortz} Lortz R, Meingast C, Rykov A I and Tajima S 2003 \textit{Phys.
Rev. Lett.} \textbf{91} 207001

\bibitem{nho} Nho K and Manousakis E 2001 \textit{Phys. Rev. B} \textbf{64} 144513

\bibitem{coleman} Coleman M and Lipa J A 1995 \textit{Phys. Rev. Lett.} \textbf{74}
286

\bibitem{gasparini} Gasparini F M, Kimball M O and Mooney K P 2001 \textit{J.
Phys.: Condens. Matter} \textbf{13} 4871

\bibitem{lyard} Lyard L \textit{et al}. 2002 \textit{Phys. Rev B} \textbf{66} 180502(R)

\bibitem{karpinski} Karpinski J \textit{et al}. 2003 \textit{Supercond. Sci. Technol.}
\textbf{16} 221

\bibitem{angst} Angst M, Puzniak R, Wisniewski A, Jun J, Kazakov S M, Karpinski J, Roos J and Keller H 2002 \textit{Phys. Rev. Lett.} \textbf{88} 167004

\bibitem{panagopoulos} Panagopoulos C, Rainford B D, Xiang T, Scott C A, Kambara M and Inoue I H, 2001 \textit{Phys. Rev. B} \textbf{64} 094514

\bibitem{dicastro} Di Castro D, Khasanov R, Grimaldi C, Karpinski J, Kazakov S M, Br\"{u}tsch R and Keller H 2005 \textit{Phys. Rev. B} \textbf{72} 094504

\end{thebibliography}
\end{document}